\documentclass[twocolumn,10pt,aps,prl,nofootinbib]{revtex4-1}

\usepackage{graphicx}	
\usepackage{color}
\usepackage{units}

\usepackage[colorlinks=true,linkcolor=red,citecolor=blue,urlcolor=green]{hyperref}

\begin{document}

%
%

\title{Active narrowband filtering, line narrowing and gain using ladder electromagnetically induced transparency in an optically thick atomic vapour}

\author{J. Keaveney$^{1}$}
\author{A. Sargsyan$^{2}$}
\author{D. Sarkisyan$^{2}$}
\author{A. Papoyan$^{2}$}
\author{C. S. Adams$^{1}$}
\affiliation{$^{1}$Joint Quantum Centre (JQC) Durham-Newcastle, Department of Physics, Durham University, South Road, Durham, DH1 3LE, United Kingdom}
\affiliation{$^{2}$Institute for Physical Research, National Academy of Sciences - Ashtarak 2, 0203, Armenia}

\begin{abstract}
Electromagnetically induced transparency (EIT) resonances using the $5\rm{S}_{1/2}\rightarrow5\rm{P}_{3/2}\rightarrow5\rm{D}_{5/2}$ ladder-system in optically thick Rb atomic vapour are studied. 
We observe a strong line narrowing effect and gain at the $5\rm{S}_{1/2}\rightarrow5\rm{P}_{3/2}$ transition wavelength due to an energy-pooling assisted frequency conversion with characteristics similar to four-wave mixing. As a result it is possible to observe tunable and switchable transparency resonances with amplitude close to 100\% and a linewidth of 15~MHz. In addition, the large line narrowing effect allows resolution of $^{85}$Rb~$5\rm{D}_{5/2}$ hyperfine structure even in the presence of strong power broadening.
\end{abstract}


\maketitle


\section{Introduction}

Coherently prepared atomic media are finding an ever increasing range of applications including high resolution metrology~\cite{Knappe2004}, 
precision magnetometry~\cite{Shah2007}, 
quantum memory~\cite{Hammerer2010}, 
entanglement~\cite{Boyer2008} and 
single-photon sources~\cite{MacRae2012,
Dudin2012,
Peyronel2012,
Maxwell2013}. 
For some applications it is necessary to post filter the output to remove the control light. 
For this reason, and other potential applications such as remote sensing~\cite{Rudolf2012}, 
atomic narrowband filtering has evolved as an active topic of research~\cite{Abel2009,Zielinska2012}. 
In addition, many of the above applications require a high optical depth. Interestingly a high optical depth can result in a line narrowing effect~\cite{Lukin1997} 
that could be useful for filtering applications. However, if the high optical depth is associated with high atomic density, other effects such as the Lorentz shift~\cite{Keaveney2012} 
and broadening~\cite{Weller2011a} 
also become important.

\begin{figure}[b]
\includegraphics[width=0.49\textwidth,angle=0]{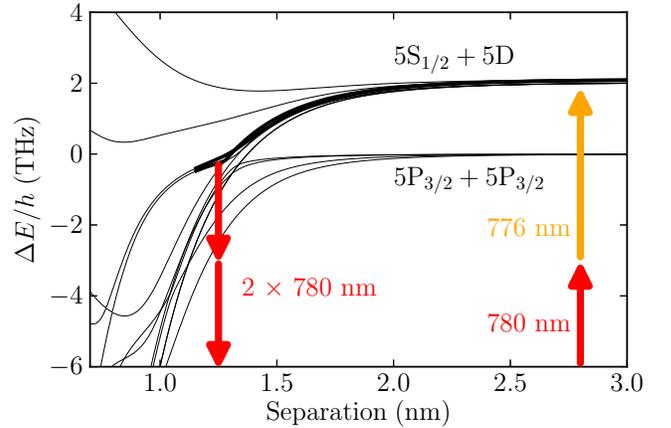}
\caption{Schematic showing how the energy pooling process can be responsible for gain in the system. Lines are energies in the pair state basis relative to the energy of two 5P$_{3/2}$ atoms at infinite separation. Initially an atom is excited to the 5D manifold by excitation with 780 and 776~nm photons, whilst a second atom remains in its ground state (pair state 5S$_{1/2}$,5D$_{5/2}$). During a collision the pair-state energies are modified, following the path in bold and allowing the transition to 5P$_{3/2}$,5P$_{3/2}$. Both of these 5P$_{3/2}$ atoms can decay with the emission of two 780~nm photons.}
\label{fig:energypooling}
\end{figure}

In this paper we investigate ladder electromagnetically-induced transparency (EIT) system based on the $5\rm{S}_{1/2}\rightarrow5\rm{P}_{3/2}\rightarrow5\rm{D}_{5/2}$ two-photon transition in Rb vapour.
Although this ladder system has been studied extensively in previous work~\cite{Gea-Banacloche1995, 
Moseley1995, 
Jin1995, 
Badger2001, 
Sargsyan2010a, 
Noh2012, 
Moon2013}, 
here we focus on novel effects that occur at high optical depths and high coupling laser Rabi frequency. In particular we show that the line narrowing effect due to high optical depth can be more efficient in a ladder-type system because, in contrast to the $\Lambda$-system studied previously~\cite{Lukin1997}, the coupling laser frequency is far detuned from the the ground-state transition and thus experiences much less absorption. In addition we find that at high density and high coupling Rabi frequency it is possible to observe gain due to a frequency conversion process analogous to that observed in $\Lambda$-systems~\cite{Li1996a, 
Mccormick2007}.
In a ladder system, however, the process is modified by energy pooling, allowing wavelength conversion over a larger energy scale. We illustrate this point in figure~\ref{fig:energypooling}. The use of a weak probe beam means that most of the atomic population remains in the ground state, while the strong coupling beam further promotes the atoms that are excited by the probe into the 5D manifold. The energy pooling process requires a collision between two atoms, so it is best to think of a joint two-atom system, where the atoms before collision are in the pair state 5S$_{1/2}$5D$_{5/2}$. This state has more energy than the nearby joint state 5P$_{3/2}$5P$_{3/2}$, but by less than the thermal energy $k_{B}T$, so in a collision the joint state can be transferred into 5P$_{3/2}$5P$_{3/2}$. Since the atoms are now both in 5P$_{3/2}$, they can decay with the emission of two 780 nm photons producing gain in the probe beam. This four-wave mixing is phase-matched using a co-propagating coupling beam. However, EIT requires a counter-propagating control beam to have a near Doppler-free resonance, so both co- and counter-propagating beams are required.

While it is possible to achieve near-100\% transparency by increasing the coupling laser power, in systems without active gain this comes at the cost of power broadening of the EIT feature, eventually leading to an Autler-Townes split doublet \cite{Anisimov2011}. In our case the gain process creates the high transmission without a significant increase in the linewidth.
%
The combination of line narrowing and gain results in a tunable narrowband filter with contrast (the amount of probe transmission, relative to the input probe intensity) of 100\% and a transparency width of order 15~MHz (FWHM). In addition, we show how density-induced line narrowing allows high resolution spectroscopy of the excited-state hyperfine structure~\cite{Noh2012} even in the presence of strong power broadening.

The paper is arranged as follows: In Section 2 we describe the experimental setup. In section 3 we present the results, and finally in section 4 we summarise the main findings.

\section{Experimental details}

\begin{figure}[t]
\includegraphics[width=0.49\textwidth,angle=0]{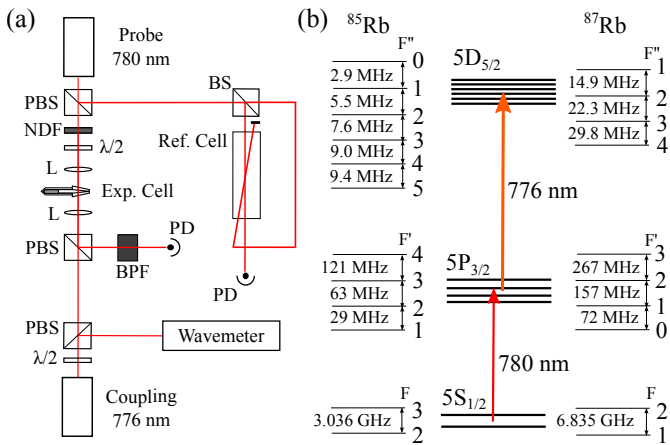}
\caption{(a) Experimental setup and (b) energy level scheme showing hyperfine splittings of each state for the two Rb isotopes. PBS - polarising beam splitter; $\lambda/2$ - half-wave plate; L - lens; PD - photodiode; BS - 50:50 beam splitter; NDF - neutral density filter; BPF - narrowband interference filters (see main text). The frequency of the coupling beam is monitored using a wavemeter. The probe laser frequency is scanned around the D2 resonance frequency, and is calibrated using pump-probe spectroscopy in a reference cell in the same way as refs~\cite{Siddons2008b,Keaveney2012,Keaveney2012a}.}
\label{fig:setup}
\end{figure}

\begin{figure*}[t]
\includegraphics[width=0.8\textwidth,angle=0]{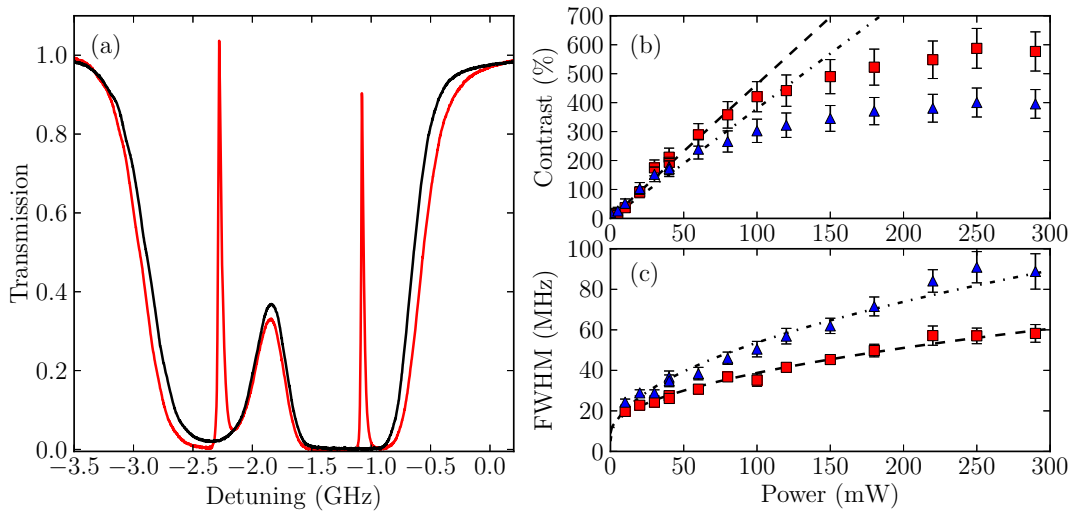}
\caption{Dependence of EIT features on the coupling laser power. Probe power 4~$\mu$W, vapour temperature 95$^{\circ}$C. (a) Example data showing the active filtering process. Black - coupling laser off. Red - 20~mW. (b) Contrast and (c) full-width at half-maximum (FWHM) of the EIT feature as the coupling laser power is varied. The contrast is initially linear with power (dashed black line) before saturating at high power. The linewidth increases with the square root of the power (black dashed line), in agreement with equation~\ref{eq:OD}. For both panels (b) and (c), red squares indicate the EIT feature around the $^{85}$Rb~F$=3\rightarrow F'=2,3,4$ transitions ($\sim-1.2$~GHz detuning), while blue triangles indicate the feature centred around the $^{87}$Rb~F$=2\rightarrow F'=1,2,3$ transitions ($\sim-2.4$~GHz detuning). Zero on the detuning axis represents the weighted center of the D2 line.}
\label{fig:power}
\end{figure*}

The experimental set-up and level scheme are shown in Fig.~\ref{fig:setup}(a) and (b), respectively. 
A probe beam (wavelength 780.2~nm) is tuned around resonance with the $5\rm{S}_{1/2}\rightarrow5\rm{P}_{3/2}$ manifolds in Rb. A counter-propagating coupling beam (wavelength 775.9~nm) is tuned on resonance with the $5\rm{P}_{3/2}\rightarrow5\rm{D}_{5/2}$ transition. The energy levels of Rb relevant to ladder EIT using the $5\rm{S}_{1/2}\rightarrow5\rm{P}_{3/2}\rightarrow5\rm{D}_{5/2}$ two-photon transition are shown in Fig.~\ref{fig:setup}(b). Both beams are derived from external cavity diode lasers with linewidths less than 1~MHz. The probe beam has a power of $\sim1~\mu$W in order to avoid saturating the first excitation step, while the coupling laser has a power up to 300~mW. Both signal and control light are focused to a spot size with a $1/{\rm e}^{2}$ radius of 27~$\mu$m at the centre of an uncoated 4 mm-long vapour cell containing natural Rb with no buffer gas. The Rayleigh range of the focussed beams, 1.5~mm, is comparable to the cell length. The probe (signal) beam is detected on a high gain photodiode. The probe and coupling beams have linear orthogonal polarizations. A polarizing beam splitting cube is used to remove reflected coupling light from the path of the detector. In addition, 3 narrowband interference filters that each transmit 93\% at 780~nm and 0.01\% at 776~nm are used to remove any residual control light. The extinction of the control light to a level that is insignificant was checked by adding an additional interference filter. 
The Rb vapour pressure, $N$, is varied between 10$^{11}$~cm$^{-3}$ up to 2$\times$10$^{14}$~cm$^{-3}$ by changing the cell temperature between 50 and 165$^\circ$C. It is important to note that the cell windows and Rb reservoir are heated pseudo-independently, and in the experiments presented here we held the window temperature constant, while varying the reservoir temperature.

The relatively small spot size and short cell length allows control laser intensities in excess of 1~kW~cm$^{-2}$ to be attained. This is 4 orders of magnitude higher than used in other high resolution experiments on ladder EIT in Rb~\cite{Moon2013}. The cell windows are fabricated from polished sapphire windows with parallel faces, which are cut perpendicular to the C-axis to minimise birefringence as discussed in refs~\cite{Sarkisyan2009book,Sargsyan2008}. However, there is still some small residual birefringence. 
The cell is aligned such that the back reflection from the cell window is perfectly overlapped with the interaction region. To realise both Doppler-free EIT and four-wave mixing requires both co- and counter-propagating control beams. In our experiment this is provided by the reflections in the cell. A single sapphire window can reflect up to 26\% of the beam if the interference between the front and back surfaces is constructive. However, this back reflection can be reduced to 0.1\% by temperature tuning the window~\cite{Jahier2000}. A strong back reflection of the control light is important to the amplification effects observed below. Whilst in principle the gain could be optimised by the use of, say, an external mirror to enhance the contribution of back reflections, this is far from a trivial process this mirror must reflect the coupling beam (776 nm), while being transparent for the probe radiation (780 nm), as well as requiring careful consideration of mode-matching conditions. The fact that the probe and coupling wavelengths are similar allows the energy pooling process to operate (and thus produce the gain), but makes independent control of the beams non-trivial.

\section{Results}

A typical transmission spectra as the probe laser is scanned through resonance is shown in Fig.~\ref{fig:power}(a). Two Doppler-broadened absorption lines are observed corresponding to the $F=2\rightarrow F'$ and $F=3\rightarrow F'$ transitions on the Rb D2 line for $^{87}$Rb and $^{85}$Rb, respectively. For reference we also show the spectra with the control laser off. Without the coupling laser the transmission on resonance for the $^{85}$Rb line is less than 0.1\%. When the coupling laser is turned on, EIT features with very high contrast are observed. With coupling power of  20 mW, the peak transmission increases to $>$90\%, and has a full-width at half-maximum (FWHM) of just 20~MHz. The positions of the transparency features correspond to the two-photon resonances, consequently the position of the transparency peak can be changed by tuning the control laser.

\begin{figure*}[t]
\includegraphics[width=0.8\textwidth,angle=0]{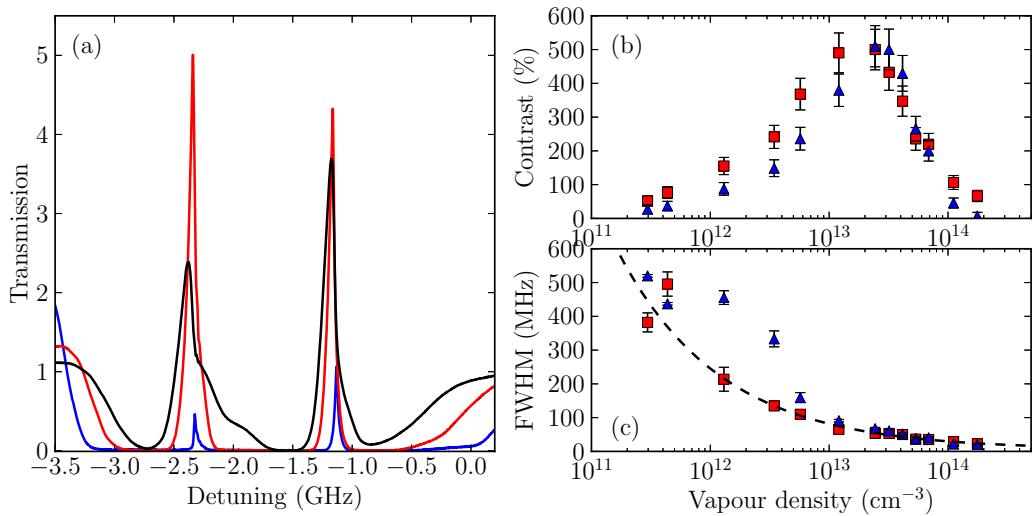}
\caption{Dependence of EIT features on the vapour density. Probe power 6~$\mu$W, coupling power 250~mW. (a) Example data showing the probe transmission as the probe detuning is varied across the D2 resonance line at various temperatures. Black - 99$^{\circ}$C. Red - 130$^{\circ}$C. Blue - 150$^{\circ}$C. (b) Contrast and (c) linewidth of the EIT feature with increasing atomic density. The linewidth decreases with density proportional to $1/\sqrt{\textrm{OD}}$, in accordance with Eq.~\ref{eq:OD}. For both panels (b) and (c), red squares indicate the EIT feature around the $^{85}$Rb~F$=3\rightarrow F'=2,3,4$ transitions, while blue triangles indicate the feature centred around the $^{87}$Rb~F$=2\rightarrow F'=1,2,3$ transitions.}
\label{fig:density}
\end{figure*}

For low coupling power, the contrast between the transmission without and with the control laser is linearly proportion to the control laser power as shown in Fig.~\ref{fig:power}(b). Surprisingly, for powers greater than of order 20~mW the transmission is greater than 100\%. This enhanced transmission indicates gain. For high powers the gain saturates at a value that depends strongly on the optical depth (see below). We find that the gain feature is only observed if the cell is aligned such that the back reflection of the control light overlaps with the interaction region and the window temperature is close to the region of maximal reflection. To observe gain requires that control photons at 776~nm are converted into signal photons at 780~nm. We propose that this effect occurs via the combination of EIT and energy-pooling. The transparency induced by the counter-propagating coupling beam suppresses absorption, while energy pooling enables frequency conversion of the retro-reflected 776~nm coupling light into 780~nm photons. 
Consequently, as for blue (420~nm) light generation from the decay of the 6P$_{1/2,3/2}$ levels~\cite{Moseley1995, 
Zibrov2002a, 
Meijer2006, 
Akulshin2009, 
Vernier2010}, 
it is likely that the efficiency is enhanced by the energy pooling process~\cite{Weller2013a}. 
If the cell is misaligned slightly then we observe 780~nm~photons in a direction that is no longer parallel to the signal beam axis, in a way that is commensurate with momentum conservation in the 4-wave mixing process. In this case the detector may be moved off axis such that only the gain features are observed without the background of the transmitted signal light. In Fig.~\ref{fig:power}(c) we show the width of the transparency feature as a function of laser power. The width is found to be proportional to the square root of the laser power.
For an ideal 3-level system a linear dependence is expected, however, Doppler broadening and optical pumping leads to such a non-linear dependence~\cite{Ye2002}. 
 
The height and width of the transmission resonances also depend strongly on the atomic density. In Fig.~\ref{fig:density} we characterize the density dependence.
In Fig.~\ref{fig:density}(a) we show the spectrum at three different cell temperatures, corresponding to three atomic densities. The coupling laser is stabilised using a wavemeter, however the wavemeter itself may have drifted a little over the course of the experiment, and this is responsible for the slight change in the peak gain position. Such a small drift does not affect the magnitude of the gain significantly. We observe the largest gain at a density of order 10$^{13}$~cm$^{-3}$ (temperature around 120$^{\circ}$C). At higher density the gain is reduced, see Fig.~\ref{fig:density}(b).
The collisional processes involved in energy pooling are strongly dependent on the density, and initially increase the amount of gain we observe. However this tails off at high density, which we attribute to dipole-dipole interactions, which enhance the energy pooling process. Instead of just 5p5p being populated, higher energy pair states can be formed~\cite{Weller2013a} due to the enhanced interaction energy between the atoms, and this decreases the amount of gain at 780nm.
The bandwidth of the filter is also reduced, as shown in Fig.~\ref{fig:density}(c), which is evidence of the line narrowing effect in EIT at high optical depth and is well known~\cite{Lukin1997}. The narrowing can be characterized by a simple formula for the the linewidth of a dark resonance in the optically thick limit, 
\begin{equation}
\gamma_{\rm EIT} =\frac{\Omega_{\rm c}^2}{\sqrt{\gamma_{12}\gamma_{2}}}\frac{1}{\sqrt{\rm OD}}~,
\label{eq:OD}
\end{equation}
where ${\rm OD} = N\sigma \ell$ is the optical depth with the optical cross-section $\sigma=3\lambda^2/2\pi$ for a closed two-level system, $\gamma_{2}$ and $\gamma_{12}$ are the decay rates of intermediate state population and the coherence between intermediate and ground states, respectively.
Thus the dark resonance linewidth scales inversely with the square root of the optical depth, and can become substantially smaller than the single-atom power-broadened linewidth given by the coupling Rabi frequency $\Omega_{\rm c}$. In Fig.~\ref{fig:density}(c) we show that the linewidth accurately follows the predicted $1/\sqrt{\textrm{OD}}$ behaviour. We attribute the large hump in the $^{87}$Rb data (blue triangles) to the presence of the $^{85}$Rb resonance in close proximity; the spectral features at the detuning associated with the $^{87}$Rb  resonance at low density also depend on the shape of the wings of the $^{85}$Rb resonance. This becomes less of an issue at higher density when both lines are optically thick, and the data return to the $1/\sqrt{\textrm{OD}}$ behaviour as expected.

\begin{figure}[t]
\includegraphics[width=0.46\textwidth,angle=0]{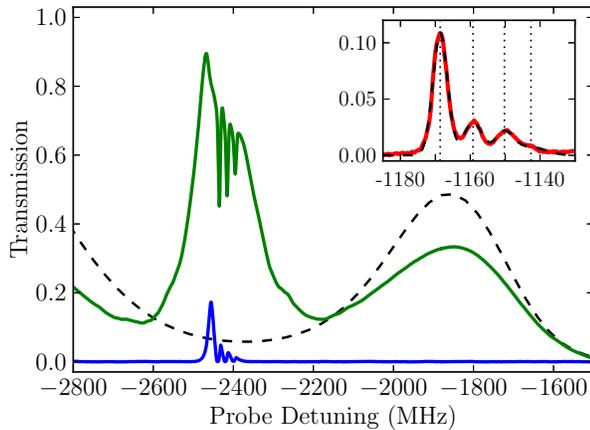}
\caption{Resolution of the hyperfine splitting of the $^{87}$Rb 5D$_{5/2}$ manifold. Green - 95C. Blue - 145C. The black dashed line shows a probe-only spectrum at 90$^{\circ}$C. The inset shows the d-state hyperfine structure of $^{85}$Rb (red), and the black dashed line shows a fit to four Gaussian functions with the correct hyperfine splittings (dotted vertical lines).}
\label{fig:dstatehyperfine}
\end{figure}

Finally, in order to illustrate how this line narrowing effect can be exploited we show in Fig.~\ref{fig:dstatehyperfine} a spectrum where the hyperfine splitting of the 5D$_{5/2}$ state is resolved even though the coupling laser Rabi frequency is much larger than the hyperfine splitting. Although ladder EIT allows high resolution spectroscopy at moderate power~\cite{Moon2013}, it is still remarkable that such high resolution can be obtained when the coupling laser intensity is four orders of magnitude larger. Further work is needed to understand the details of these spectra in the high density regime.

\section{Summary}

In summary, we have demonstrated an active tunable narrowband atomic filter with a contrast that can be greater than 100\% whilst simultaneously having a bandwidth of less than 100~MHz. The filter exploits both line narrowing and gain in optically thick Rb vapor.  Such a filter could have advantages compared with more conventional Fabry-Perot etalons in that the filter is rapidly switchable and tunable, and because there is no loss. The switching speed is related to the linewidth of the feature, which constrains the minimum pulse width. Taking a conservative estimate, assuming an EIT linewidth of 10~MHz, a Fourier-transform limited Gaussian pulse profile would need a pulse width of $>100$~ns.
However, for applications in quantum optics further work is needed to explore whether the inherent gain adds significant noise. Similar effects may be possible using other transitions as well as different atomic media (Cs, K, Na, etc.) allowing development over a range of useful wavelengths. We have also shown that even for highly intense lasers in excess of $\sim$1~kW/cm$^{2}$ the formation of narrow dark resonances allows high resolution spectroscopy of excited states, with resolution only limited by the laser stability.

The authors would like to thank Christophe Vaillant for the calculation of the two-atom potentials presented in figure 1. JK and CSA acknowledge financial support from EPSRC and Durham University. AS and DS acknowledge support from the State Science Committee MES RA, in the frame of research project no. SCS13-1C029.

\section*{References}

\bibliographystyle{apsrev4-1}
\bibliography{library}

\begin{thebibliography}{10}%
\makeatletter
\providecommand \@ifxundefined [1]{%
 \ifx #1\undefined \expandafter \@firstoftwo
 \else \expandafter \@secondoftwo
\fi
}%
\providecommand \@ifnum [1]{%
 \ifnum #1\expandafter \@firstoftwo
 \else \expandafter \@secondoftwo
\fi
}%
\providecommand \enquote [1]{``#1''}%
\providecommand \bibnamefont  [1]{#1}%
\providecommand \bibfnamefont [1]{#1}%
\providecommand \citenamefont [1]{#1}%
\providecommand\href[0]{\@sanitize\@href}%
\providecommand\@href[1]{\endgroup\@@startlink{#1}\endgroup\@@href}%
\providecommand\@@href[1]{#1\@@endlink}%
\providecommand \@sanitize [0]{\begingroup\catcode`\&12\catcode`\#12\relax}%
\@ifxundefined \pdfoutput {\@firstoftwo}{%
 \@ifnum{\z@=\pdfoutput}{\@firstoftwo}{\@secondoftwo}%
}{%
 \providecommand\@@startlink[1]{\leavevmode\special{html:<a href="#1">}}%
 \providecommand\@@endlink[0]{\special{html:</a>}}%
}{%
 \providecommand\@@startlink[1]{%
  \leavevmode
  \pdfstartlink
   attr{/Border[0 0 1 ]/H/I/C[0 1 1]}%
   user{/Subtype/Link/A<</Type/Action/S/URI/URI(#1)>>}%
  \relax
 }%
 \providecommand\@@endlink[0]{\pdfendlink}%
}%
\providecommand \url  [0]{\begingroup\@sanitize \@url }%
\providecommand \@url [1]{\endgroup\@href {#1}{\urlprefix}}%
\providecommand \urlprefix [0]{URL }%
\providecommand \Eprint[0]{\href }%
\@ifxundefined \urlstyle {%
  \providecommand \doi [1]{doi:\discretionary{}{}{}#1}%
}{%
  \providecommand \doi [0]{doi:\discretionary{}{}{}\begingroup
  \urlstyle{rm}\Url }%
}%
\providecommand \doibase [0]{http://dx.doi.org/}%
\providecommand \Doi[1]{\href{\doibase#1}}%
\providecommand \bibAnnote [3]{%
  \BibitemShut{#1}%
  \begin{quotation}\noindent
    \textsc{Key:}\ #2\\\textsc{Annotation:}\ #3%
  \end{quotation}%
}%
\providecommand \bibAnnoteFile [2]{%
  \IfFileExists{#2}{\bibAnnote {#1} {#2} {\input{#2}}}{}%
}%
\providecommand \typeout [0]{\immediate \write \m@ne }%
\providecommand \selectlanguage [0]{\@gobble}%
\providecommand \bibinfo [0]{\@secondoftwo}%
\providecommand \bibfield [0]{\@secondoftwo}%
\providecommand \translation [1]{[#1]}%
\providecommand \BibitemOpen[0]{}%
\providecommand \bibitemStop [0]{}%
\providecommand \bibitemNoStop [0]{.\EOS\space}%
\providecommand \EOS [0]{\spacefactor3000\relax}%
\providecommand \BibitemShut [1]{\csname bibitem#1\endcsname}%
\bibitem{Knappe2004}%
  \BibitemOpen
  \bibfield{author}{%
  \bibinfo {author} {\bibfnamefont{S.}~\bibnamefont{Knappe}}, \bibinfo {author}
  {\bibfnamefont{V.}~\bibnamefont{Shah}}, \bibinfo {author}
  {\bibfnamefont{P.~D.~D.}\ \bibnamefont{Schwindt}}, \bibinfo {author}
  {\bibfnamefont{L.}~\bibnamefont{Hollberg}}, \bibinfo {author}
  {\bibfnamefont{J.}~\bibnamefont{Kitching}}, \bibinfo {author}
  {\bibfnamefont{L.-A.}\ \bibnamefont{Liew}},\ and\ \bibinfo {author}
  {\bibfnamefont{J.}~\bibnamefont{Moreland}},\ }%
  \bibfield{journal}{%
  \Doi{10.1063/1.1787942}{\bibinfo {journal} {Appl. Phys. Lett.}}\ }%
  \textbf{\bibinfo {volume} {85}},\ \bibinfo {pages} {1460} (\bibinfo {year}
  {2004}),\ ISSN \bibinfo {issn} {00036951},\
  \url{http://link.aip.org/link/APPLAB/v85/i9/p1460/s1&Agg=doi}%
  \bibAnnoteFile{NoStop}{Knappe2004}%
\bibitem{Shah2007}%
  \BibitemOpen
  \bibfield{author}{%
  \bibinfo {author} {\bibfnamefont{V.}~\bibnamefont{Shah}}, \bibinfo {author}
  {\bibfnamefont{S.}~\bibnamefont{Knappe}}, \bibinfo {author}
  {\bibfnamefont{P.~D.~D.}\ \bibnamefont{Schwindt}},\ and\ \bibinfo {author}
  {\bibfnamefont{J.}~\bibnamefont{Kitching}},\ }%
  \bibfield{journal}{%
  \Doi{10.1038/nphoton.2007.201}{\bibinfo {journal} {Nature Photon.}}\ }%
  \textbf{\bibinfo {volume} {1}},\ \bibinfo {pages} {649} (\bibinfo {month}
  {Nov.}\ \bibinfo {year} {2007}),\ ISSN \bibinfo {issn} {1749-4885},\
  \url{http://www.nature.com/doifinder/10.1038/nphoton.2007.201}%
  \bibAnnoteFile{NoStop}{Shah2007}%
\bibitem{Hammerer2010}%
  \BibitemOpen
  \bibfield{author}{%
  \bibinfo {author} {\bibfnamefont{K.}~\bibnamefont{Hammerer}}, \bibinfo
  {author} {\bibfnamefont{A.~S.}\ \bibnamefont{S{\o}rensen}},\ and\ \bibinfo
  {author} {\bibfnamefont{E.~S.}\ \bibnamefont{Polzik}},\ }%
  \bibfield{journal}{%
  \Doi{10.1103/RevModPhys.82.1041}{\bibinfo {journal} {Rev. Mod. Phys.}}\ }%
  \textbf{\bibinfo {volume} {82}},\ \bibinfo {pages} {1041} (\bibinfo {month}
  {Apr.}\ \bibinfo {year} {2010}),\ ISSN \bibinfo {issn} {0034-6861},\
  \url{http://link.aps.org/doi/10.1103/RevModPhys.82.1041}%
  \bibAnnoteFile{NoStop}{Hammerer2010}%
\bibitem{Boyer2008}%
  \BibitemOpen
  \bibfield{author}{%
  \bibinfo {author} {\bibfnamefont{V.}~\bibnamefont{Boyer}}, \bibinfo {author}
  {\bibfnamefont{A.~M.}\ \bibnamefont{Marino}}, \bibinfo {author}
  {\bibfnamefont{R.~C.}\ \bibnamefont{Pooser}},\ and\ \bibinfo {author}
  {\bibfnamefont{P.~D.}\ \bibnamefont{Lett}},\ }%
  \bibfield{journal}{%
  \Doi{10.1126/science.1158275}{\bibinfo {journal} {Science}}\ }%
  \textbf{\bibinfo {volume} {321}},\ \bibinfo {pages} {544} (\bibinfo {month}
  {Jul.}\ \bibinfo {year} {2008}),\ ISSN \bibinfo {issn} {1095-9203},\
  \url{http://www.ncbi.nlm.nih.gov/pubmed/18556517}%
  \bibAnnoteFile{NoStop}{Boyer2008}%
\bibitem{MacRae2012}%
  \BibitemOpen
  \bibfield{author}{%
  \bibinfo {author} {\bibfnamefont{A.}~\bibnamefont{MacRae}}, \bibinfo {author}
  {\bibfnamefont{T.}~\bibnamefont{Brannan}}, \bibinfo {author}
  {\bibfnamefont{R.}~\bibnamefont{Achal}},\ and\ \bibinfo {author}
  {\bibfnamefont{A.~I.}\ \bibnamefont{Lvovsky}},\ }%
  \bibfield{journal}{%
  \Doi{10.1103/PhysRevLett.109.033601}{\bibinfo {journal} {Phys. Rev. Lett.}}\
  }%
  \textbf{\bibinfo {volume} {109}},\ \bibinfo {pages} {033601} (\bibinfo
  {month} {Jul.}\ \bibinfo {year} {2012}),\ ISSN \bibinfo {issn} {0031-9007},\
  \url{http://link.aps.org/doi/10.1103/PhysRevLett.109.033601}%
  \bibAnnoteFile{NoStop}{MacRae2012}%
\bibitem{Dudin2012}%
  \BibitemOpen
  \bibfield{author}{%
  \bibinfo {author} {\bibfnamefont{Y.~O.}\ \bibnamefont{Dudin}}\ and\ \bibinfo
  {author} {\bibfnamefont{A.}~\bibnamefont{Kuzmich}},\ }%
  \bibfield{journal}{%
  \Doi{10.1126/science.1217901}{\bibinfo {journal} {Science}}\ }%
  \textbf{\bibinfo {volume} {336}},\ \bibinfo {pages} {887} (\bibinfo {month}
  {May}\ \bibinfo {year} {2012}),\ ISSN \bibinfo {issn} {1095-9203},\
  \url{http://www.ncbi.nlm.nih.gov/pubmed/22517325}%
  \bibAnnoteFile{NoStop}{Dudin2012}%
\bibitem{Peyronel2012}%
  \BibitemOpen
  \bibfield{author}{%
  \bibinfo {author} {\bibfnamefont{T.}~\bibnamefont{Peyronel}}, \bibinfo
  {author} {\bibfnamefont{O.}~\bibnamefont{Firstenberg}}, \bibinfo {author}
  {\bibfnamefont{Q.-Y.}\ \bibnamefont{Liang}}, \bibinfo {author}
  {\bibfnamefont{S.}~\bibnamefont{Hofferberth}}, \bibinfo {author}
  {\bibfnamefont{A.~V.}\ \bibnamefont{Gorshkov}}, \bibinfo {author}
  {\bibfnamefont{T.}~\bibnamefont{Pohl}}, \bibinfo {author}
  {\bibfnamefont{M.~D.}\ \bibnamefont{Lukin}},\ and\ \bibinfo {author}
  {\bibfnamefont{V.}~\bibnamefont{Vuleti\'{c}}},\ }%
  \bibfield{journal}{%
  \Doi{10.1038/nature11361}{\bibinfo {journal} {Nature}}\ }%
  \textbf{\bibinfo {volume} {488}},\ \bibinfo {pages} {57} (\bibinfo {month}
  {Aug.}\ \bibinfo {year} {2012}),\ ISSN \bibinfo {issn} {1476-4687},\
  \url{http://www.ncbi.nlm.nih.gov/pubmed/22832584}%
  \bibAnnoteFile{NoStop}{Peyronel2012}%
\bibitem{Maxwell2013}%
  \BibitemOpen
  \bibfield{author}{%
  \bibinfo {author} {\bibfnamefont{D.}~\bibnamefont{Maxwell}}, \bibinfo
  {author} {\bibfnamefont{D.~J.}\ \bibnamefont{Szwer}}, \bibinfo {author}
  {\bibfnamefont{D.}~\bibnamefont{Paredes-Barato}}, \bibinfo {author}
  {\bibfnamefont{H.}~\bibnamefont{Busche}}, \bibinfo {author}
  {\bibfnamefont{J.~D.}\ \bibnamefont{Pritchard}}, \bibinfo {author}
  {\bibfnamefont{A.}~\bibnamefont{Gauguet}}, \bibinfo {author}
  {\bibfnamefont{K.~J.}\ \bibnamefont{Weatherill}}, \bibinfo {author}
  {\bibfnamefont{M.~P.~A.}\ \bibnamefont{Jones}},\ and\ \bibinfo {author}
  {\bibfnamefont{C.~S.}\ \bibnamefont{Adams}},\ }%
  \bibfield{journal}{%
  \Doi{10.1103/PhysRevLett.110.103001}{\bibinfo {journal} {Phys. Rev. Lett.}}\
  }%
  \textbf{\bibinfo {volume} {110}},\ \bibinfo {pages} {103001} (\bibinfo
  {month} {Mar.}\ \bibinfo {year} {2013}),\ ISSN \bibinfo {issn} {0031-9007},\
  \url{http://link.aps.org/doi/10.1103/PhysRevLett.110.103001}%
  \bibAnnoteFile{NoStop}{Maxwell2013}%
\bibitem{Rudolf2012}%
  \BibitemOpen
  \bibfield{author}{%
  \bibinfo {author} {\bibfnamefont{A.}~\bibnamefont{Rudolf}}\ and\ \bibinfo
  {author} {\bibfnamefont{T.}~\bibnamefont{Walther}},\ }%
  \bibfield{journal}{%
  \bibinfo {journal} {Opt. Lett.}\ }%
  \textbf{\bibinfo {volume} {37}},\ \bibinfo {pages} {4477} (\bibinfo {month}
  {Nov.}\ \bibinfo {year} {2012}),\ ISSN \bibinfo {issn} {1539-4794},\
  \url{http://www.ncbi.nlm.nih.gov/pubmed/23114335}%
  \bibAnnoteFile{NoStop}{Rudolf2012}%
\bibitem{Abel2009}%
  \BibitemOpen
  \bibfield{author}{%
  \bibinfo {author} {\bibfnamefont{R.~P.}\ \bibnamefont{Abel}}, \bibinfo
  {author} {\bibfnamefont{A.~K.}\ \bibnamefont{Mohapatra}}, \bibinfo {author}
  {\bibfnamefont{M.~G.}\ \bibnamefont{Bason}}, \bibinfo {author}
  {\bibfnamefont{J.~D.}\ \bibnamefont{Pritchard}}, \bibinfo {author}
  {\bibfnamefont{K.~J.}\ \bibnamefont{Weatherill}}, \bibinfo {author}
  {\bibfnamefont{U.}~\bibnamefont{Raitzsch}},\ and\ \bibinfo {author}
  {\bibfnamefont{C.~S.}\ \bibnamefont{Adams}},\ }%
  \bibfield{journal}{%
  \Doi{10.1063/1.3086305}{\bibinfo {journal} {Appl. Phys. Lett.}}\ }%
  \textbf{\bibinfo {volume} {94}},\ \bibinfo {pages} {071107} (\bibinfo {year}
  {2009}),\ ISSN \bibinfo {issn} {00036951},\
  \url{http://link.aip.org/link/APPLAB/v94/i7/p071107/s1&Agg=doi}%
  \bibAnnoteFile{NoStop}{Abel2009}%
\bibitem{Zielinska2012}%
  \BibitemOpen
  \bibfield{author}{%
  \bibinfo {author} {\bibfnamefont{J.~A.}\ \bibnamefont{Zielińska}}, \bibinfo
  {author} {\bibfnamefont{F.~A.}\ \bibnamefont{Beduini}}, \bibinfo {author}
  {\bibfnamefont{N.}~\bibnamefont{Godbout}},\ and\ \bibinfo {author}
  {\bibfnamefont{M.~W.}\ \bibnamefont{Mitchell}},\ }%
  \bibfield{journal}{%
  \bibinfo {journal} {Opt. Lett.}\ }%
  \textbf{\bibinfo {volume} {37}},\ \bibinfo {pages} {524} (\bibinfo {month}
  {Feb.}\ \bibinfo {year} {2012}),\ ISSN \bibinfo {issn} {1539-4794},\
  \url{http://www.ncbi.nlm.nih.gov/pubmed/22344094}%
  \bibAnnoteFile{NoStop}{Zielinska2012}%
\bibitem{Lukin1997}%
  \BibitemOpen
  \bibfield{author}{%
  \bibinfo {author} {\bibfnamefont{M.~D.}\ \bibnamefont{Lukin}}, \bibinfo
  {author} {\bibfnamefont{M.}~\bibnamefont{Fleischhauer}}, \bibinfo {author}
  {\bibfnamefont{A.~S.}\ \bibnamefont{Zibrov}}, \bibinfo {author}
  {\bibfnamefont{H.~G.}\ \bibnamefont{Robinson}}, \bibinfo {author}
  {\bibfnamefont{V.~L.}\ \bibnamefont{Velichansky}}, \bibinfo {author}
  {\bibfnamefont{L.}~\bibnamefont{Hollberg}}, \bibinfo {author}
  {\bibfnamefont{M.~O.}\ \bibnamefont{Scully}},\ and\ \bibinfo {author}
  {\bibfnamefont{S.}~\bibnamefont{Raman}},\ }%
  \bibfield{journal}{%
  \bibinfo {journal} {Phys. Rev. Lett.}\ }%
  \textbf{\bibinfo {volume} {79}} (\bibinfo {year} {1997})%
  \bibAnnoteFile{NoStop}{Lukin1997}%
\bibitem{Keaveney2012}%
  \BibitemOpen
  \bibfield{author}{%
  \bibinfo {author} {\bibfnamefont{J.}~\bibnamefont{Keaveney}}, \bibinfo
  {author} {\bibfnamefont{A.}~\bibnamefont{Sargsyan}}, \bibinfo {author}
  {\bibfnamefont{U.}~\bibnamefont{Krohn}}, \bibinfo {author}
  {\bibfnamefont{I.~G.}\ \bibnamefont{Hughes}}, \bibinfo {author}
  {\bibfnamefont{D.}~\bibnamefont{Sarkisyan}},\ and\ \bibinfo {author}
  {\bibfnamefont{C.~S.}\ \bibnamefont{Adams}},\ }%
  \bibfield{journal}{%
  \Doi{10.1103/PhysRevLett.108.173601}{\bibinfo {journal} {Phys. Rev. Lett.}}\
  }%
  \textbf{\bibinfo {volume} {108}},\ \bibinfo {pages} {173601} (\bibinfo
  {month} {Apr.}\ \bibinfo {year} {2012}),\ ISSN \bibinfo {issn} {0031-9007},\
  \url{http://link.aps.org/doi/10.1103/PhysRevLett.108.173601}%
  \bibAnnoteFile{NoStop}{Keaveney2012}%
\bibitem{Weller2011a}%
  \BibitemOpen
  \bibfield{author}{%
  \bibinfo {author} {\bibfnamefont{L.}~\bibnamefont{Weller}}, \bibinfo {author}
  {\bibfnamefont{R.~J.}\ \bibnamefont{Bettles}}, \bibinfo {author}
  {\bibfnamefont{P.}~\bibnamefont{Siddons}}, \bibinfo {author}
  {\bibfnamefont{C.~S.}\ \bibnamefont{Adams}},\ and\ \bibinfo {author}
  {\bibfnamefont{I.~G.}\ \bibnamefont{Hughes}},\ }%
  \bibfield{journal}{%
  \Doi{10.1088/0953-4075/44/19/195006}{\bibinfo {journal} {J. Phys. B}}\ }%
  \textbf{\bibinfo {volume} {44}},\ \bibinfo {pages} {195006} (\bibinfo {month}
  {Oct.}\ \bibinfo {year} {2011}),\ ISSN \bibinfo {issn} {0953-4075},\
  \url{http://stacks.iop.org/0953-4075/44/i=19/a=195006?key=crossref.6053c04e063ad80a26ce7648067c84ae}%
  \bibAnnoteFile{NoStop}{Weller2011a}%
\bibitem{Gea-Banacloche1995}%
  \BibitemOpen
  \bibfield{author}{%
  \bibinfo {author} {\bibfnamefont{J.}~\bibnamefont{Gea-Banacloche}}, \bibinfo
  {author} {\bibfnamefont{Y.-Q.}\ \bibnamefont{Li}}, \bibinfo {author}
  {\bibfnamefont{S.-Z.}\ \bibnamefont{Jin}},\ and\ \bibinfo {author}
  {\bibfnamefont{M.}~\bibnamefont{Xiao}},\ }%
  \bibfield{journal}{%
  \Doi{10.1103/PhysRevA.51.576}{\bibinfo {journal} {Phys. Rev. A}}\ }%
  \textbf{\bibinfo {volume} {51}},\ \bibinfo {pages} {576} (\bibinfo {month}
  {Jan.}\ \bibinfo {year} {1995}),\ ISSN \bibinfo {issn} {1050-2947},\
  \url{http://link.aps.org/doi/10.1103/PhysRevA.51.576}%
  \bibAnnoteFile{NoStop}{Gea-Banacloche1995}%
\bibitem{Moseley1995}%
  \BibitemOpen
  \bibfield{author}{%
  \bibinfo {author} {\bibfnamefont{R.~R.}\ \bibnamefont{Moseley}}, \bibinfo
  {author} {\bibfnamefont{S.}~\bibnamefont{Shepherd}}, \bibinfo {author}
  {\bibfnamefont{D.~J.}\ \bibnamefont{Fulton}}, \bibinfo {author}
  {\bibfnamefont{B.~D.}\ \bibnamefont{Sinclair}},\ and\ \bibinfo {author}
  {\bibfnamefont{M.~H.}\ \bibnamefont{Dunn}},\ }%
  \bibfield{journal}{%
  \Doi{10.1016/0030-4018(95)00316-Z}{\bibinfo {journal} {Opt. Commun.}}\ }%
  \textbf{\bibinfo {volume} {119}},\ \bibinfo {pages} {61} (\bibinfo {month}
  {Aug.}\ \bibinfo {year} {1995}),\ ISSN \bibinfo {issn} {00304018},\
  \url{http://linkinghub.elsevier.com/retrieve/pii/003040189500316Z}%
  \bibAnnoteFile{NoStop}{Moseley1995}%
\bibitem{Jin1995}%
  \BibitemOpen
  \bibfield{author}{%
  \bibinfo {author} {\bibfnamefont{S.}~\bibnamefont{Jin}}, \bibinfo {author}
  {\bibfnamefont{Y.}~\bibnamefont{Li}},\ and\ \bibinfo {author}
  {\bibfnamefont{M.}~\bibnamefont{Xiao}},\ }%
  \bibfield{journal}{%
  \Doi{10.1016/0030-4018(95)00362-C}{\bibinfo {journal} {Opt. Commun.}}\ }%
  \textbf{\bibinfo {volume} {119}},\ \bibinfo {pages} {90} (\bibinfo {month}
  {Aug.}\ \bibinfo {year} {1995}),\ ISSN \bibinfo {issn} {00304018},\
  \url{http://linkinghub.elsevier.com/retrieve/pii/003040189500362C}%
  \bibAnnoteFile{NoStop}{Jin1995}%
\bibitem{Badger2001}%
  \BibitemOpen
  \bibfield{author}{%
  \bibinfo {author} {\bibfnamefont{S.~D.}\ \bibnamefont{Badger}}, \bibinfo
  {author} {\bibfnamefont{I.~G.}\ \bibnamefont{Hughes}},\ and\ \bibinfo
  {author} {\bibfnamefont{C.~S.}\ \bibnamefont{Adams}},\ }%
  \bibfield{journal}{%
  \Doi{10.1088/0953-4075/34/22/107}{\bibinfo {journal} {J. Phys. B}}\ }%
  \textbf{\bibinfo {volume} {34}},\ \bibinfo {pages} {L749} (\bibinfo {month}
  {Nov.}\ \bibinfo {year} {2001}),\ ISSN \bibinfo {issn} {0953-4075},\
  \url{http://stacks.iop.org/0953-4075/34/i=22/a=107?key=crossref.b34de0a3532176086c4e07d64bb72d03}%
  \bibAnnoteFile{NoStop}{Badger2001}%
\bibitem{Sargsyan2010a}%
  \BibitemOpen
  \bibfield{author}{%
  \bibinfo {author} {\bibfnamefont{A.}~\bibnamefont{Sargsyan}}, \bibinfo
  {author} {\bibfnamefont{D.}~\bibnamefont{Sarkisyan}}, \bibinfo {author}
  {\bibfnamefont{U.}~\bibnamefont{Krohn}}, \bibinfo {author}
  {\bibfnamefont{J.}~\bibnamefont{Keaveney}},\ and\ \bibinfo {author}
  {\bibfnamefont{C.~S.}\ \bibnamefont{Adams}},\ }%
  \bibfield{journal}{%
  \Doi{10.1103/PhysRevA.82.045806}{\bibinfo {journal} {Phys. Rev. A}}\ }%
  \textbf{\bibinfo {volume} {82}},\ \bibinfo {pages} {045806} (\bibinfo {year}
  {2010}),\ \url{http://link.aps.org/doi/10.1103/PhysRevA.82.045806}%
  \bibAnnoteFile{NoStop}{Sargsyan2010a}%
\bibitem{Noh2012}%
  \BibitemOpen
  \bibfield{author}{%
  \bibinfo {author} {\bibfnamefont{H.-R.}\ \bibnamefont{Noh}}\ and\ \bibinfo
  {author} {\bibfnamefont{H.~S.}\ \bibnamefont{Moon}},\ }%
  \bibfield{journal}{%
  \Doi{10.1103/PhysRevA.85.033817}{\bibinfo {journal} {Phys. Rev. A}}\ }%
  \textbf{\bibinfo {volume} {85}},\ \bibinfo {pages} {033817} (\bibinfo {month}
  {Mar.}\ \bibinfo {year} {2012}),\ ISSN \bibinfo {issn} {1050-2947},\
  \url{http://link.aps.org/doi/10.1103/PhysRevA.85.033817}%
  \bibAnnoteFile{NoStop}{Noh2012}%
\bibitem{Moon2013}%
  \BibitemOpen
  \bibfield{author}{%
  \bibinfo {author} {\bibfnamefont{H.~S.}\ \bibnamefont{Moon}}\ and\ \bibinfo
  {author} {\bibfnamefont{H.-R.}\ \bibnamefont{Noh}},\ }%
  \bibfield{journal}{%
  \bibinfo {journal} {Opt. Express}\ }%
  \textbf{\bibinfo {volume} {21}},\ \bibinfo {pages} {7447} (\bibinfo {month}
  {Mar.}\ \bibinfo {year} {2013}),\ ISSN \bibinfo {issn} {1094-4087},\
  \url{http://www.ncbi.nlm.nih.gov/pubmed/23546128}%
  \bibAnnoteFile{NoStop}{Moon2013}%
\bibitem{Li1996a}%
  \BibitemOpen
  \bibfield{author}{%
  \bibinfo {author} {\bibfnamefont{Y.~Q.}\ \bibnamefont{Li}}\ and\ \bibinfo
  {author} {\bibfnamefont{M.}~\bibnamefont{Xiao}},\ }%
  \bibfield{journal}{%
  \bibinfo {journal} {Opt. Lett.}\ }%
  \textbf{\bibinfo {volume} {21}},\ \bibinfo {pages} {1064} (\bibinfo {month}
  {Jul.}\ \bibinfo {year} {1996}),\ ISSN \bibinfo {issn} {0146-9592},\
  \url{http://www.ncbi.nlm.nih.gov/pubmed/19876253}%
  \bibAnnoteFile{NoStop}{Li1996a}%
\bibitem{Mccormick2007}%
  \BibitemOpen
  \bibfield{author}{%
  \bibinfo {author} {\bibfnamefont{C.~F.}\ \bibnamefont{Mccormick}}, \bibinfo
  {author} {\bibfnamefont{V.}~\bibnamefont{Boyer}}, \bibinfo {author}
  {\bibfnamefont{E.}~\bibnamefont{Arimondo}},\ and\ \bibinfo {author}
  {\bibfnamefont{P.~D.}\ \bibnamefont{Lett}},\ }%
  \bibfield{journal}{%
  \bibinfo {journal} {Opt. Lett.}\ }%
  \textbf{\bibinfo {volume} {32}},\ \bibinfo {pages} {2006} (\bibinfo {year}
  {2007})%
  \bibAnnoteFile{NoStop}{Mccormick2007}%
\bibitem{Anisimov2011}%
  \BibitemOpen
  \bibfield{author}{%
  \bibinfo {author} {\bibfnamefont{P.}~\bibnamefont{Anisimov}}, \bibinfo
  {author} {\bibfnamefont{J.}~\bibnamefont{Dowling}},\ and\ \bibinfo {author}
  {\bibfnamefont{B.}~\bibnamefont{Sanders}},\ }%
  \bibfield{journal}{%
  \Doi{10.1103/PhysRevLett.107.163604}{\bibinfo {journal} {Phys. Rev. Lett.}}\
  }%
  \textbf{\bibinfo {volume} {107}},\ \bibinfo {pages} {163604} (\bibinfo
  {month} {Oct.}\ \bibinfo {year} {2011}),\ ISSN \bibinfo {issn} {0031-9007},\
  \url{http://link.aps.org/doi/10.1103/PhysRevLett.107.163604}%
  \bibAnnoteFile{NoStop}{Anisimov2011}%
\bibitem{Siddons2008b}%
  \BibitemOpen
  \bibfield{author}{%
  \bibinfo {author} {\bibfnamefont{P.}~\bibnamefont{Siddons}}, \bibinfo
  {author} {\bibfnamefont{C.~S.}\ \bibnamefont{Adams}}, \bibinfo {author}
  {\bibfnamefont{C.}~\bibnamefont{Ge}},\ and\ \bibinfo {author}
  {\bibfnamefont{I.~G.}\ \bibnamefont{Hughes}},\ }%
  \bibfield{journal}{%
  \Doi{10.1088/0953-4075/41/15/155004}{\bibinfo {journal} {J. Phys. B}}\ }%
  \textbf{\bibinfo {volume} {41}},\ \bibinfo {pages} {155004} (\bibinfo {month}
  {Aug.}\ \bibinfo {year} {2008}),\ ISSN \bibinfo {issn} {0953-4075},\
  \url{http://stacks.iop.org/0953-4075/41/i=15/a=155004?key=crossref.00c7138be904263053df130d39a795f7}%
  \bibAnnoteFile{NoStop}{Siddons2008b}%
\bibitem{Keaveney2012a}%
  \BibitemOpen
  \bibfield{author}{%
  \bibinfo {author} {\bibfnamefont{J.}~\bibnamefont{Keaveney}}, \bibinfo
  {author} {\bibfnamefont{I.~G.}\ \bibnamefont{Hughes}}, \bibinfo {author}
  {\bibfnamefont{A.}~\bibnamefont{Sargsyan}}, \bibinfo {author}
  {\bibfnamefont{D.}~\bibnamefont{Sarkisyan}},\ and\ \bibinfo {author}
  {\bibfnamefont{C.~S.}\ \bibnamefont{Adams}},\ }%
  \bibfield{journal}{%
  \Doi{10.1103/PhysRevLett.109.233001}{\bibinfo {journal} {Phys. Rev. Lett.}}\
  }%
  \textbf{\bibinfo {volume} {109}},\ \bibinfo {pages} {233001} (\bibinfo
  {month} {Dec.}\ \bibinfo {year} {2012}),\ ISSN \bibinfo {issn} {0031-9007},\
  \url{http://link.aps.org/doi/10.1103/PhysRevLett.109.233001}%
  \bibAnnoteFile{NoStop}{Keaveney2012a}%
\bibitem{Sarkisyan2009book}%
  \BibitemOpen
  \bibfield{author}{%
  \bibinfo {author} {\bibfnamefont{D.}~\bibnamefont{Sarkisyan}}\ and\ \bibinfo
  {author} {\bibfnamefont{A.}~\bibnamefont{Papoyan}}\ }%
  (\bibinfo {publisher} {Nova Science},\ \bibinfo {year} {2009})\
  Chap.~\bibinfo {chapter} {3},\ ISBN \bibinfo {isbn} {978-1-60741- 025-6}%
  \bibAnnoteFile{NoStop}{Sarkisyan2009book}%
\bibitem{Sargsyan2008}%
  \BibitemOpen
  \bibfield{author}{%
  \bibinfo {author} {\bibfnamefont{A.}~\bibnamefont{Sargsyan}}, \bibinfo
  {author} {\bibfnamefont{G.}~\bibnamefont{Hakhumyan}}, \bibinfo {author}
  {\bibfnamefont{A.}~\bibnamefont{Papoyan}}, \bibinfo {author}
  {\bibfnamefont{D.}~\bibnamefont{Sarkisyan}}, \bibinfo {author}
  {\bibfnamefont{A.}~\bibnamefont{Atvars}},\ and\ \bibinfo {author}
  {\bibfnamefont{M.}~\bibnamefont{Auzinsh}},\ }%
  \bibfield{journal}{%
  \Doi{10.1063/1.2960346}{\bibinfo {journal} {Appl. Phys. Lett.}}\ }%
  \textbf{\bibinfo {volume} {93}},\ \bibinfo {pages} {021119} (\bibinfo {year}
  {2008}),\ ISSN \bibinfo {issn} {00036951},\
  \url{http://link.aip.org/link/APPLAB/v93/i2/p021119/s1&Agg=doi}%
  \bibAnnoteFile{NoStop}{Sargsyan2008}%
\bibitem{Jahier2000}%
  \BibitemOpen
  \bibfield{author}{%
  \bibinfo {author} {\bibfnamefont{E.}~\bibnamefont{Jahier}}, \bibinfo {author}
  {\bibfnamefont{J.}~\bibnamefont{Gu\'{e}na}}, \bibinfo {author}
  {\bibfnamefont{P.}~\bibnamefont{Jacquier}}, \bibinfo {author}
  {\bibfnamefont{M.}~\bibnamefont{Lintz}}, \bibinfo {author}
  {\bibfnamefont{A.~V.}\ \bibnamefont{Papoyan}},\ and\ \bibinfo {author}
  {\bibfnamefont{M.~A.}\ \bibnamefont{Bouchiat}},\ }%
  \bibfield{journal}{%
  \Doi{10.1007/s003400000388}{\bibinfo {journal} {Appl. Phys. B}}\ }%
  \textbf{\bibinfo {volume} {71}},\ \bibinfo {pages} {561} (\bibinfo {month}
  {Oct.}\ \bibinfo {year} {2000}),\ ISSN \bibinfo {issn} {0946-2171},\
  \url{http://www.springerlink.com/openurl.asp?genre=article&id=doi:10.1007/s003400000388}%
  \bibAnnoteFile{NoStop}{Jahier2000}%
\bibitem{Zibrov2002a}%
  \BibitemOpen
  \bibfield{author}{%
  \bibinfo {author} {\bibfnamefont{A.~S.}\ \bibnamefont{Zibrov}}, \bibinfo
  {author} {\bibfnamefont{M.}~\bibnamefont{Lukin}}, \bibinfo {author}
  {\bibfnamefont{L.}~\bibnamefont{Hollberg}},\ and\ \bibinfo {author}
  {\bibfnamefont{M.~O.}\ \bibnamefont{Scully}},\ }%
  \bibfield{journal}{%
  \Doi{10.1103/PhysRevA.65.051801}{\bibinfo {journal} {Phys. Rev. A}}\ }%
  \textbf{\bibinfo {volume} {65}},\ \bibinfo {pages} {051801(R)} (\bibinfo
  {month} {Apr.}\ \bibinfo {year} {2002}),\ ISSN \bibinfo {issn} {1050-2947},\
  \url{http://link.aps.org/doi/10.1103/PhysRevA.65.051801}%
  \bibAnnoteFile{NoStop}{Zibrov2002a}%
\bibitem{Meijer2006}%
  \BibitemOpen
  \bibfield{author}{%
  \bibinfo {author} {\bibfnamefont{T.}~\bibnamefont{Meijer}}, \bibinfo {author}
  {\bibfnamefont{J.~D.}\ \bibnamefont{White}}, \bibinfo {author}
  {\bibfnamefont{B.}~\bibnamefont{Smeets}}, \bibinfo {author}
  {\bibfnamefont{M.}~\bibnamefont{Jeppesen}},\ and\ \bibinfo {author}
  {\bibfnamefont{R.~E.}\ \bibnamefont{Scholten}},\ }%
  \bibfield{journal}{%
  \bibinfo {journal} {Opt. Lett.}\ }%
  \textbf{\bibinfo {volume} {31}},\ \bibinfo {pages} {1002} (\bibinfo {month}
  {Apr.}\ \bibinfo {year} {2006}),\ ISSN \bibinfo {issn} {0146-9592},\
  \url{http://www.ncbi.nlm.nih.gov/pubmed/16599237}%
  \bibAnnoteFile{NoStop}{Meijer2006}%
\bibitem{Akulshin2009}%
  \BibitemOpen
  \bibfield{author}{%
  \bibinfo {author} {\bibfnamefont{A.~M.}\ \bibnamefont{Akulshin}}, \bibinfo
  {author} {\bibfnamefont{R.~J.}\ \bibnamefont{McLean}}, \bibinfo {author}
  {\bibfnamefont{A.~I.}\ \bibnamefont{Sidorov}},\ and\ \bibinfo {author}
  {\bibfnamefont{P.}~\bibnamefont{Hannaford}},\ }%
  \bibfield{journal}{%
  \bibinfo {journal} {Opt. Express}\ }%
  \textbf{\bibinfo {volume} {17}},\ \bibinfo {pages} {22861} (\bibinfo {month}
  {Dec.}\ \bibinfo {year} {2009}),\ ISSN \bibinfo {issn} {1094-4087},\
  \url{http://www.ncbi.nlm.nih.gov/pubmed/20052212}%
  \bibAnnoteFile{NoStop}{Akulshin2009}%
\bibitem{Vernier2010}%
  \BibitemOpen
  \bibfield{author}{%
  \bibinfo {author} {\bibfnamefont{A.}~\bibnamefont{Vernier}}, \bibinfo
  {author} {\bibfnamefont{S.}~\bibnamefont{Franke-Arnold}}, \bibinfo {author}
  {\bibfnamefont{E.}~\bibnamefont{Riis}},\ and\ \bibinfo {author}
  {\bibfnamefont{A.~S.}\ \bibnamefont{Arnold}},\ }%
  \bibfield{journal}{%
  \bibinfo {journal} {Opt. Express}\ }%
  \textbf{\bibinfo {volume} {18}},\ \bibinfo {pages} {17020} (\bibinfo {month}
  {Aug.}\ \bibinfo {year} {2010}),\ ISSN \bibinfo {issn} {1094-4087},\
  \url{http://www.ncbi.nlm.nih.gov/pubmed/20721090}%
  \bibAnnoteFile{NoStop}{Vernier2010}%
\bibitem{Weller2013a}%
  \BibitemOpen
  \bibfield{author}{%
  \bibinfo {author} {\bibfnamefont{L.}~\bibnamefont{Weller}}, \bibinfo {author}
  {\bibfnamefont{R.~J.}\ \bibnamefont{Bettles}}, \bibinfo {author}
  {\bibfnamefont{C.~L.}\ \bibnamefont{Vaillant}}, \bibinfo {author}
  {\bibfnamefont{M.~A.}\ \bibnamefont{Zentile}}, \bibinfo {author}
  {\bibfnamefont{R.~M.}\ \bibnamefont{Potvliege}}, \bibinfo {author}
  {\bibfnamefont{C.~S.}\ \bibnamefont{Adams}},\ and\ \bibinfo {author}
  {\bibfnamefont{I.~G.}\ \bibnamefont{Hughes}},\ }%
  \bibfield{journal}{%
  \bibinfo {journal} {arXiv.org},\ \bibinfo {pages} {1308.0129}}%
   (\bibinfo {year} {2013}),\
  \Eprint{http://arxiv.org/abs/arXiv:1308.0129v1}{arXiv:arXiv:1308.0129v1}%
  \bibAnnoteFile{NoStop}{Weller2013a}%
\bibitem{Ye2002}%
  \BibitemOpen
  \bibfield{author}{%
  \bibinfo {author} {\bibfnamefont{C.}~\bibnamefont{Ye}}\ and\ \bibinfo
  {author} {\bibfnamefont{A.}~\bibnamefont{Zibrov}},\ }%
  \bibfield{journal}{%
  \Doi{10.1103/PhysRevA.65.023806}{\bibinfo {journal} {Phys. Rev. A}}\ }%
  \textbf{\bibinfo {volume} {65}},\ \bibinfo {pages} {023806} (\bibinfo {month}
  {Jan.}\ \bibinfo {year} {2002}),\ ISSN \bibinfo {issn} {1050-2947},\
  \url{http://link.aps.org/doi/10.1103/PhysRevA.65.023806}%
  \bibAnnoteFile{NoStop}{Ye2002}%
\end{thebibliography}%

\end{document}